\PassOptionsToPackage{dvipsnames}{xcolor}
\PassOptionsToPackage{oneside}{geometry}
\PassOptionsToPackage{pdfencoding=auto}{hyperref}
\PassOptionsToPackage{mathlines}{lineno}
\documentclass[twocolumn, twocolappendix, oneside]{aastex631}
\received{\today}
\shorttitle{\cogsworth}
\shortauthors{Wagg, Breivik, Renzo \& Price-Whelan}
\graphicspath{{figures/}}

\usepackage{lipsum}
\usepackage{physics}
\usepackage{multirow}
\usepackage{xspace}
\usepackage{natbib}
\usepackage{fontawesome5}
\usepackage{wrapfig}
\usepackage[figuresright]{rotating}
\usepackage[frozencache,cachedir=minted]{minted}

\usepackage[hang,flushmargin]{footmisc} 

\usepackage{styles/todo_notes}
\usepackage{styles/abbreviations}
\usepackage{styles/docs_icons}

\makeatletter
\newcommand{\unit}[1]{%
    \,\mathrm{#1}\checknextarg}
\newcommand{\checknextarg}{\@ifnextchar\bgroup{\gobblenextarg}{}}
\newcommand{\gobblenextarg}[1]{\,\mathrm{#1}\@ifnextchar\bgroup{\gobblenextarg}{}}
\makeatother

\newif\ifstartedinmathmode
\newcommand{\msun}{%
  \relax\ifmmode\startedinmathmodetrue\else\startedinmathmodefalse\fi
  {\ifstartedinmathmode\unit{M_{\odot}}\else$\unit{M_{\odot}}$\fi}\xspace%
}

\newif\ifstartedinmathmode
\newcommand{\rsun}{%
  \relax\ifmmode\startedinmathmodetrue\else\startedinmathmodefalse\fi
  {\ifstartedinmathmode\unit{R_{\odot}}\else$\unit{R_{\odot}}$\fi}\xspace%
}

\definecolor{codecolour}{HTML}{905cc4}

\newcommand{\codeLink}[2]{{\href{https://cogsworth.readthedocs.io/en/latest/api/cogsworth.#2.#1.html}{\color{codecolour} \texttt{#1}}}}
\newcommand{\codestyle}[1]{{\color{codecolour} \texttt{#1}}}

\newcommand{\tutorialIcon}{{\color{codecolour}{\faLaptopCode}}}
\newcommand{\tutorialIconLink}[1]{\href{#1}{\tutorialIcon}}
\newcommand{\tutorialLink}[2]{\href{#1}{{\color{codecolour}#2}}}

\newcommand{\invisibleedit}[1]{#1}

\begin{document}


\title{\cogsworth: A \gala of \cosmic proportions combining binary stellar evolution and galactic dynamics}


\newcommand{\UW}{\affiliation{Department of Astronomy, University of Washington, Seattle, WA, 98195}}
\newcommand{\cca}{\affiliation{Center for Computational Astrophysics, Flatiron Institute, 162 Fifth Ave, New York, NY, 10010, USA}}
\newcommand{\cmu}{\affiliation{McWilliams Center for Cosmology and Astrophysics, Department of Physics, Carnegie Mellon University, Pittsburgh, PA 15213, USA}}
\newcommand{\arizona}{\affiliation{University of Arizona, Department of Astronomy \& Steward Observatory, 933 N. Cherry Ave., Tucson, AZ 85721, USA}}

\author[0000-0001-6147-5761]{Tom Wagg}
\UW{}
\cca{}

\author[0000-0001-5228-6598]{Katelyn Breivik}
\cmu{}

\author[0000-0002-6718-9472]{Mathieu Renzo}
\arizona{}

\author[0000-0003-0872-7098]{Adrian~M.~Price-Whelan}
\cca{}


\correspondingauthor{Tom Wagg}
\email{tomjwagg@gmail.com}

\begin{abstract}

    We present \cogsworth, an open-source Python tool for producing self-consistent population synthesis and galactic dynamics simulations. \cogsworth allows users to (1) sample a population of binaries and star formation history, (2) perform rapid (binary) stellar evolution, (3) integrate orbits through the galaxy, and (4) inspect the full evolutionary history of each star or compact object, along with their positions and kinematics. It supports post-processing hydrodynamical zoom-in simulations for more realistic galactic potentials and star formation histories, accounting for initial spatial stellar clustering and complex potentials. Alternatively, several analytic models are available for galactic potentials and star formation histories. \cogsworth can also transform the intrinsic simulated population into an observed population using dust maps, bolometric correction functions, and survey selection functions.
    
    We provide a detailed explanation of the functionality of \cogsworth and demonstrate its capabilities through a series of use cases: (1) We predict the spatial distribution of compact objects and runaways in both dwarf and Milky-Way-like galaxies, (2) using a star cluster from a hydrodynamical simulation, we show how supernovae can change the orbits of stars in several ways, and (3) we predict the separation of disrupted binary stellar companions on the sky and create a synthetic \gaia colour-magnitude diagram. \invisibleedit{We use \cogsworth to demonstrate that both binarity and the galactic potential have a significant impact of the present-day positions of massive stars.} We designed \cogsworth and its online documentation to provide a powerful tool for constraining binary evolution, but also a flexible and accessible resource for the entire community.
\end{abstract}




\section{Introduction}

The majority of stars are born in binaries and multiple star systems \citep[e.g.,][]{Kroupa+1995:1995MNRAS.277.1491K,Duchene+2013:2013ARA&A..51..269D,Moe+2017,Oliva+2020:2020A&A...644A..41O,Offner+2023:2023ASPC..534..275O}, a large subset of which will exchange mass at some point in their lifetime \citep[e.g][]{Podsiadlowski+1992:1992ApJ...391..246P,Sana+2012,deMink+2014}. These massive stars play a critical role in the formation and evolution of galaxies as a result of their feedback \citep[e.g.,][]{Dekel+1986:1986ApJ...303...39D,Hopkins+2012:2012MNRAS.421.3522H,Nomoto+2013:2013ARA&A..51..457N,Somerville+2015:2015ARA&A..53...51S,Naab+2017:2017ARA&A..55...59N}.

However, binary evolution remains uncertain, with many parameters such as common-envelope efficiency, mass transfer efficiency, angular momentum loss due to mass transfer and the mean magnitude of supernova natal kicks unconstrained across several orders of magnitude \citep[e.g.,][]{Janka+2012:2012ARNPS..62..407J,Ivanova+2013, Katsuda+2018,Ivanova+2020:2020cee..book.....I,Ropke+2023:2023LRCA....9....2R, Marchant2023}.

Single massive stars are not expected to migrate far from their birth location before reaching core-collapse due to their short lifetimes \citep[$\lesssim50$\,Myr, e.g., ][]{zapartas:17}. However, binary stars may disrupt after an initial supernova event, ejecting the secondary star from the system with its orbital velocity \citep[e.g.,][]{Blaauw+1961,Eldridge+2011:2011MNRAS.414.3501E,Renzo+2019:2019A&A...624A..66R}. Thus, close massive binaries that disrupt can lead to the displacement of secondary stars significantly farther from star-forming regions. The present-day positions and kinematics of massive stars and binary products are therefore strongly impacted by changes in binary physics that alter the separation prior to supernova. This means that comparing simulations of positions and kinematics of stars and compact objects to observations will enable constraints on binary stellar evolution parameters.

The use of positions and kinematics as tracers of binary evolution has been considered in the past. Recent work has shown the importance of accounting for the galactic potential, which can change the velocity of kicked objects \citep[e.g.,][]{Disberg+2024:2024A&A...687A.272D}. It is also important to consider the inclination or timing of a supernova kick relative to the galactic orbit, since, for example, a kick out of the galactic plane at an object's highest galactic vertical position will have a strong effect on its final position. Failing to consider impacts from both a galactic potential and kicks (i.e. velocity impulses) will lead to misleading conclusions regarding the final spatial distributions of the population. Some studies have considered using the Galactic potential at the present-day position of objects to place a lower limit on the peculiar velocity at birth and constrain supernova kicks \citep{Repetto+2012:2012MNRAS.425.2799R, Repetto+2015:2015MNRAS.453.3341R, Repetto+2017:2017MNRAS.467..298R, atri:19}, but the accuracy of this method is debated \citep{Mandel+2016:2016MNRAS.456..578M}. Other work has considered the impact of the Galactic potential for individual special cases, rather than at a population level. For example, \citet{Evans+2020:2020MNRAS.497.5344E} considered the orbits of hyper-runaway candidates evolving through the Milky Way potential, whilst \citet{Neuhauser+2020:2020MNRAS.498..899N} developed software for tracing the motion of stars to investigate the recent nearby supernovae that ejected $\zeta$ Ophiuchi. \invisibleedit{Additionally, \citet{Andrews+2022:2022ApJ...930..159A} considered galactic orbits of synthetic populations to place constraints on black hole natal kicks based on observations of a microlensed black hole. Previous work has also worked towards connecting population synthesis and galactic dynamics. Through a combination of the \texttt{COMPAS} population synthesis code \citep{COMPAS} and \texttt{NIGO} galactic orbit integration tool \citep{Rossi+2015:2015A&C....12...11R} to make predictions for binary neutron stars, black hole neutron star binaries and pulsars \citep{Chattopadhyay+2020:2020MNRAS.494.1587C, Chattopadhyay+2021:2021MNRAS.504.3682C, Song+2024:2024arXiv240611428S}. Moreover, the \texttt{BiPoS1} code accounts for the effect of dynamical encounters on internal binary orbits and can be used to synthetise a galactic stellar population \citep{Dabringhausen+2022:2022MNRAS.510..413D}.}



In this paper we present \cogsworth, a new open-source tool for self-consistent population synthesis and galactic dynamics simulations. \cogsworth provides the theoretical infrastructure for making predictions for the positions and kinematics of massive stars and compact objects, placing these systems in the context of their host galaxy and its gravitational potential. The code is applicable to a wide range of binary products, both common and rare, from walkaway and runaway stars to X-ray binaries, as well as gravitational-wave and gamma-ray burst progenitors.

The paper is structured as follows: in Section~\ref{sec:cogsworth} we explain the functionality of \cogsworth and describe its primary features and capabilities. We demonstrate these capabilities in a series of example use cases in Section~\ref{sec:use_cases}. We use \cogsworth to predict the spatial distribution of compact objects and runaways in both dwarf and Milky-Way-like galaxies. Using a cluster from a hydrodynamical simulation, we show how supernovae can change the orbits of stars in several ways. We predict the separation of disrupted binary stellar companions on the sky, as well as create a synthetic \gaia colour-magnitude diagram. In Section~\ref{sec:limitations}, we discuss the current limitations of the package and we outline planned additional future developments in Section~\ref{sec:future}.

\section{cogsworth}\label{sec:cogsworth}

\cogsworth is a code that combines binary population synthesis simulations (via \cosmic, \citealp{COSMIC}) with galactic dynamics (via \gala, \citealp{Gala}) to self-consistently use stellar and orbital evolution to rapidly derive present day positions, kinematics and demographics for complete populations of binary stars and their descendants. 

Our code is fully open-source and openly-developed (\footnoteLink{available on GitHub}{https://github.com/TomWagg/cogsworth}), pip installable (\texttt{pip install cogsworth}) and \href{https://zenodo.org/records/13709381}{indexed on Zenodo}. In this paper we describe v2.0.0 of the code. We wrote \cogsworth in Python to make it convenient and accessible, but its core dependencies are written in Fortran and C (via \cosmic and \gala respectively) for efficiency. We use automated testing via a detailed suite of unit tests, with full code coverage. Additionally, we have written thorough documentation of \cogsworth, including  $\sim$20 tutorials covering full usage of the code, several short examples and a series of longer in-depth case studies, all of which is \footnoteLink{available online}{https://cogsworth.readthedocs.io}.

\begin{figure}
    \centering
    \includegraphics[width=\columnwidth]{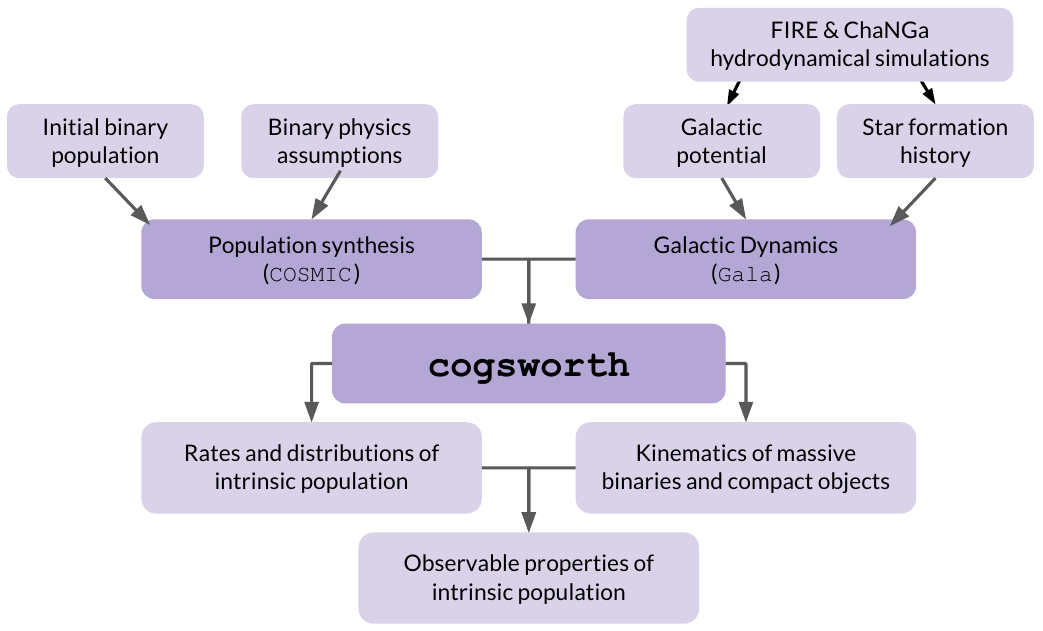}
    \caption{\cogsworth combines population synthesis and galactic dynamics self-consistently to produce rates, distributions, observables and kinematics of massive binaries and compact objects. A schematic of the input options and outputs of \cogsworth simulations.}
    \label{fig:cogsworth_overview}
\end{figure}

We describe the specific capabilities in the following subsections, and illustrate an overview of the code in Figure~\ref{fig:cogsworth_overview}. The first subsections focus on core functionality of \cogsworth, which is accessed via a \codeLink{Population}{pop}, with which one can:

\begin{itemize}
    \item [\S\ref{sec:galactic_SFH}] Sample initial galactic positions, velocities, birth times and metallicities from a star formation history model
    \item [\S\ref{sec:binary_sampling_evolution_COSMIC}] Sample and evolve a (binary) stellar population until present day
    \item [\S\ref{sec:orbit_integration}] Integrate the orbits of each binary through the galaxy, accounting for supernova kicks and disruptions
    \item [\S\ref{sec:observables}] Identify observable constituents of the present day intrinsic population
\end{itemize}
Each of these features are flexible and can be tuned to a particular use case. In addition, in \S\ref{sec:hydro_linking} we describe how one can alternatively use \cogsworth to initialise and evolve populations based on hydrodynamical zoom-in simulations. In the later subsections we detail \cogsworth's visualisation functionalities (\$\ref{sec:visuals}), details of its typical runtime (\S\ref{sec:runtime}) and data storage (\S\ref{sec:data}), and its ability to create custom citation statements for a given simulation (\S\ref{sec:citations}).

\subsection{Galactic star formation histories}\label{sec:galactic_SFH}

A galactic star formation history (SFH) defines the distributions of times, locations and metallicities at which stars are formed in a galaxy. In \cogsworth, one can flexibly adjust the SFH of a simulation with a \codeLink{StarFormationHistory}{sfh} Python class. By default, \cogsworth uses \codeLink{Wagg2022}{sfh}, an empirically motivated analytic model for the Milky Way \citep{Wagg+2022}. This model contains a low-[$\alpha$/Fe] (a.k.a.~``thin")  disc, high-[$\alpha$/Fe] (a.k.a.~``thick") disc, and bulge, each with their own spatial and temporal distributions. Metallicity is calculated as a function of birth time and galactocentric radius. This SFH accounts for effects such as the inside-out growth growth of the galaxy and radial migration \citep{Frankel+2018, Frankel+2019:2019ApJ...884...99F}. We explore this SFH further in Section~\ref{sec:sfh_use_case} and the full details of the model are given in Section 2.2.1 of \citet{Wagg+2022}.

Beyond the default SFH, we include simpler parameterised SFHs, such as \codeLink{BurstUniformDisc}{sfh}, in which stars are formed in a single burst of star formation with a fixed metallicity in a uniform disc. Additionally, we include action-based SFHs using \texttt{Agama} \citep{2019MNRAS.482.1525V}, such as \codeLink{QuasiIsothermalDisc}{sfh}, which represents a quasi-isothermal distribution function for the Milky Way disc as described in \citet{Sanders+2015:2015MNRAS.449.3479S}.

Each SFH class is designed to be modular and flexible and as such, they can be entirely customised. Users can overwrite individual distributions in a given SFH (such as changing the birth time distribution of the bulge component), or define their own entirely custom SFH. We explain how to accomplish this in \tutorialLink{https://cogsworth.readthedocs.io/en/latest/tutorials/pop_settings/initial_galaxy.html}{a tutorial}.

For each binary or single star, $i$, in a population, we use the SFH to draw the initial galactic parameters
\begin{equation}\label{eq:init_gal}
    g_{i} = \{ \tau, R, z, \phi, Z \},
\end{equation}
where $\tau$ is the lookback time (the time before the present day when the system formed), $R$ is the initial galactocentric radius, $z$ is the initial height above the plane, $\phi$ is the azimuthal angle, and $Z$ is the metallicity. For an SFH that doesn't explicitly define a distribution for galactocentric velocities, we assign the initial galactocentric velocity of a system, $i$, as follows
\begin{equation}\label{eq:init_vel}
    \va{v}_{i} = \va{v}_{\rm circ}(R_i) + \va{v}_{\rm disp},
\end{equation}
where $\va{v}_{\rm circ}$ is the circular velocity for the population's galactic potential and $\va{v}_{\rm disp}$ is an isotropic velocity dispersion, which is an input option that by default has a magnitude of $5 \unit{km}{s^{-1}}$.

\subsection{Stellar population sampling and evolution}\label{sec:binary_sampling_evolution_COSMIC}

\cogsworth uses the open-source and community-driven rapid binary population synthesis code \cosmic{} to perform the sampling of the initial (binary) stellar population and the (binary) stellar evolution \citep{COSMIC}. \cosmic{} uses fitting formulae based on single stellar tracks originally developed for the Binary Stellar Evolution (\texttt{BSE}) code \citep{Tout+1997:1997MNRAS.291..732T,pols:98,Hurley+2000:2000MNRAS.315..543H,Hurley+2002} and allows the user to rapidly sample and evolve populations of binaries with a variety of physics assumptions. With \cosmic a user can specify dynamic time resolution conditions for its outputs based on binary parameters which, for example, could be used to increase the number of outputted timesteps once a star is stripped, or during mass transfer onto a compact object to investigate X-ray binaries. One can also easily access the initial conditions of a population evolved with \cosmic, allowing for convenient reproducibility of simulations. This is additionally useful for re-running identical initial populations with alternative binary physics settings to ascertain how the evolution changes with different settings. Each of these features are directly inherited by \cogsworth.

When sampling an initial binary stellar population, a user can specify their choice of initial mass function (IMF) for drawing primary masses, a binary fraction and distributions of initial orbital period, eccentricity and mass ratio for a given \codeLink{Population}{pop} of binaries (see \citealp[Section 2.1.1 of][]{COSMIC} for all available options). Metallicities are set based on the chosen SFH model (see Section~\ref{sec:galactic_SFH}). Using these distributions, one can either draw a fixed number of systems, or specify a total mass to sample. \tutorialLink{https://cogsworth.readthedocs.io/en/latest/tutorials/pop_settings/sampling.html}{This tutorial} explains in detail how to change these settings.

\cogsworth evolves this initial binary population from their individual birth times until present day with a user-specified choice of binary physics. Any binary stellar evolution parameter that can be supplied to \cosmic can also be specified in a \codeLink{Population}{pop} in \cogsworth. These parameters cover a range of binary physics including stellar wind mass and accretion, mass transfer through Roche-lobe overflow, common-envelopes, supernova kicks, remnant mass prescriptions and tides. For a full list of the available parameters see the \href{https://cosmic-popsynth.github.io}{\cosmic documentation} and learn about changing them in \cogsworth with \tutorialLink{https://cogsworth.readthedocs.io/en/latest/tutorials/pop_settings/pop_synth.html}{this tutorial}.

\subsection{Galactic orbit integration}\label{sec:orbit_integration}

\cogsworth applies the galactic dynamics package \gala{} for the galactic orbit integration of binaries \citep{Gala}. This package allows users to integrate the orbits of sources rapidly with user-friendly functions wrapped on low-level code (primarily C) for fast computations. One can choose from numerous, flexible potentials (or even define custom potentials) through which to integrate orbits. \cogsworth uses \gala to integrate the full orbit of each binary in a population through a given galactic potential. By default, \cogsworth uses the \texttt{MilkyWayPotential2022} potential, which is fit to observations of the Milky way rotation curve, the shape of the phase-space spiral in the solar neighbourhood and a compilation of recent mass measurements of the Milky Way \citep{Eilers+2019:2019ApJ...871..120E, Darragh-Ford+2023:2023ApJ...955...74D}. \invisibleedit{\cogsworth users can apply any model for the galactic potential that is available in \gala. Each model has flexible parameters (such as mass and scale height) and can be combined into composite potentials, such as combining an NFW profile \citep{Navarro+1997:1997ApJ...490..493N} and a Plummer potential \citep{Plummer+1911:1911MNRAS..71..460P}. Additionally, one can fit an arbitrary galactic mass distribution using the self-consistent field method implemented in \gala based on \citet{Hernquist+1992:1992ApJ...386..375H} and \citet{Lowing+2011:2011MNRAS.416.2697L}. \tutorialLink{https://cogsworth.readthedocs.io/en/latest/tutorials/observables/photometry.html}{This tutorial} describes how one can change the galactic potential used in \cogsworth simulations.}

For systems that experience supernovae, \cogsworth accounts for the resulting changes in velocity. \cosmic logs the velocities imparted by Blaauw \citep{Blaauw+1961} and natal kicks \citep{Katz+1975:1975Natur.253..698K,Janka+2012:2012ARNPS..62..407J,Janka+2017:2017ApJ...837...84J} and whether a binary is disrupted by a supernova \citep[e.g.,][]{Renzo+2019:2019A&A...624A..66R}. In each case, \cogsworth transforms the resulting velocities to galactocentric coordinates (uniformly sampling a random orbital phase, $\theta$, of the binary and inclination, $\iota$, of the binary relative to the galaxy) and updating the orbit of the system. In the case of disruptions, a second orbit is produced for the secondary, tracking the binary position until the disruption and then the subsequent motion of the secondary.

Overall, this allows users to track the location of either star in a binary system at \textit{any} point in its evolutionary history. This can be used to, for instance, predict the location of supernovae or track the sites of $r$-process enrichment from binary mergers.
\subsection{Observables estimation}\label{sec:observables}

Key to applying \cogsworth{} to realistic problems and constraining our models is being able to compare simulations to observations. In this Section, we explain how users can transform intrinsic \cogsworth populations into observables.

\subsubsection{Electromagnetic observations}

We have implemented functionality to translate intrinsic stellar parameters in \cogsworth populations (such as mass, luminosity and galactic position) into observables (such as fluxes and colours). Currently, \cogsworth focuses on producing predictions for \gaia observables, but we intend to build on this with other instruments in future (see Section~\ref{sec:other-observables}). 

\cogsworth can compute the magnitude of sources in arbitrary filters by applying bolometric corrections and dust extinctions, achieved through a combination of the \texttt{dustmaps} and \texttt{isochrones} packages  \citep{2018JOSS....3..695M, Morton+2015:2015ascl.soft03010M}. \invisibleedit{We match isochrones based on the effective temperature, surface gravity and [Fe/H] metallicity of the evolved stars. We note that the MIST isochrones do not vary [$\alpha$/Fe] and as such we effectively assume a solar abundance pattern in these predictions \citep{Dotter+2016, Choi+2016:2016ApJ...823..102C}}. You can follow \tutorialLink{https://cogsworth.readthedocs.io/en/latest/tutorials/observables/photometry.html}{this tutorial} to learn how to compute photometry for \cogsworth sources. Within the Milky Way, the interplay between distance, the 3-dimensional dust distribution, and the \gaia{} scanning pattern leads to a complex selection function, but one that can be captured through the empirical selection function made available through \texttt{gaiaunlimited} \citep{Cantat-Gaudin+2023:2023A&A...669A..55C}. \cogsworth is therefore capable of predicting whether a given source (either a bound binary or star from a disrupted binary) would be detectable by \gaia. \tutorialLink{https://cogsworth.readthedocs.io/en/latest/tutorials/observables/gaia.html}{This tutorial} explains how to make predictions about the observable \gaia population.

\subsubsection{Gravitational waves}

In addition to electromagnetic observations, we consider gravitational wave detections from the inspiral of double compact objects. Stellar-mass binaries in the Milky Way will be detectable by LISA via millihertz gravitational wave emission \citep{Amaro-Seoane+2017:2017arXiv170200786A}. The \texttt{LEGWORK} package allows users to compute gravitational-wave strains and SNRs for binaries, and calculate the evolution of binary separations and eccentricities due to gravitational wave emission \citep{LEGWORK_joss, LEGWORK_apjs}. We connect \cogsworth to \texttt{LEGWORK}, allowing users to quickly calculate the LISA SNR of each binary in a population, as well as the time until its merger with a single, simple function call. \tutorialLink{https://cogsworth.readthedocs.io/en/latest/tutorials/misc/lisa.html}{This tutorial} shows an example of calculating LISA SNRs for a \cogsworth population of double white dwarfs.

\subsection{Building off hydrodynamical zoom-in simulations}\label{sec:hydro_linking}

Using an analytic model for a galactic SFH can work well for longer-lived populations, which can be expected to have erased all memory of their initial positions. However, it is unrealistic for younger stars, and in particular short-lived massive ones, which should retain significant initial spatial clustering and correlations with the surrounding ISM \citep[e.g.,][]{Sarbadhicary+2023:2023arXiv231017694S}. Such correlations are particularly important when attempting to constrain aspects of binary physics by comparing predicted present-day locations from \cogsworth models to observations of recently formed binaries.

Motivated by a need for more detailed initial spatial clustering, we include the option to initialise a \cogsworth population using hydrodynamical zoom-in simulations. These simulations are not only used to set the locations and times of star formation, but also the galactic gravitational potential. 

\subsubsection{Compatible simulations}

We currently support the post-processing for two different suites of hydrodynamical zoom-in simulations. One can use any of the public \fire simulations \citep{Wetzel+2016, Hopkins+2018:2018MNRAS.480..800H, Sanderson+2020}, which have been connected to population synthesis and cluster models successfully in the past \citep[e.g.,][]{Lamberts+2018:2018MNRAS.480.2704L, Chawla+2022:2022ApJ...931..107C, balls_of_FIRE_1, balls_of_FIRE_2, Thiele+2023:2023ApJ...945..162T}. Additionally, one can use simulations from \changa, such as the MARVEL-ous dwarfs and DC Justice League simulations \citep{Applebaum+2021:2021ApJ...906...96A, Christensen+2023:2023AAS...24140704C}. These simulation suites directly resolve the formation of giant molecular clouds and the interstellar medium (ISM), and thus capture the characteristic spatial clustering of star formation. The simulations additionally explicitly account for the feedback from stars, following predictions laid out by stellar population synthesis models - though neglecting the impact of binary evolution on the timing and positioning of supernovae (see Wagg et al.\ in prep).

\subsubsection{Snapshot preparation}\label{sec:prep_snapshots}

Hydrodynamical simulations record snapshots of the state of the simulation a specific times. \cogsworth provides a wrapper over \texttt{pynbody} \citep{pynbody} in order to prepare simulation snapshots for use as initial conditions to simulations. This functionality centres snapshots on the primary halo, using either an automatically detected halo catalogue or applying a shrinking sphere method to iteratively refine an estimate of the centre of the mass of the simulation. It additionally rotates the halo to be edge-on and then face-on, and converts data to physical units. 

\paragraph{Galactic potential} As with a regular \codeLink{Population}{pop}, before initialisation one needs to know the galactic potential and SFH of the galaxy. We provide functionality for computing a galactic potential from a simulation snapshot, accounting for stars, gas and dark matter, using the self-consistent field method implemented in \gala based on \citet{Hernquist+1992:1992ApJ...386..375H} and \citet{Lowing+2011:2011MNRAS.416.2697L}. This method fits the galactic mass distribution using a basis function expansion in spherical harmonics.

\paragraph{Initial stellar positions} The formation locations of star particles are necessary for sampling the initial positions of binary stellar populations. \cogsworth can identify these formation locations by backwards integrating the orbits of star particles through the galactic potential derived from the simulation. Note that this step is only necessary for \fire simulations, since \changa simulations store formation locations.\\

\noindent For more information on processing simulation snapshots in \cogsworth we refer interested readers to \tutorialLink{https://cogsworth.readthedocs.io/en/latest/tutorials/hydro/prep.html}{this tutorial}.

\subsubsection{Population initialisation and evolution}

A \cogsworth \ \codeLink{Population}{pop} based on the star particles and galactic potential of a hydrodynamical zoom-in is called a \codeLink{HydroPopulation}{hydro.pop}. Each star particle in a hydrodynamical simulation can represent many 100--1000s of solar masses. Given this, we use \cosmic to sample binary stellar populations from each star particle, assigning each system the same formation time and metallicity as the star particle. Based on a user-defined star particle radius, each sampled system is assigned a random position, $\va{p}_i$, from a Gaussian centred on the parent star particle such that
\begin{equation}
    \va{p}_i = \mathcal{N}(\va{p}_{{\rm sp}, i}, r),
\end{equation}
where $\va{p}_{{\rm sp}, i}$ is the position of star particle from which system $i$ was sampled and $r$ is the user choice of radius. Similarly, based on a user's choice of virial parameter, $\alpha_{\rm vir}$ (the ratio of kinetic and gravitational energy of a cluster, as defined in \citealp{Bertoldi+1992:1992ApJ...395..140B}) a velocity dispersion of each star particle is determined and used for sampling initial velocities, $\va{v}_i$, such that
\begin{equation}
    \va{v}_i = \va{v}_{{\rm sp}, i} + \va{v}_{{\rm disp}, i},
\end{equation}
where $\va{v}_{{\rm sp}, i}$ is the velocity of the star particle from which system $i$ was sampled and
\begin{equation}\label{eq:vel_disp_cluster}
    \va{v}_{{\rm disp}, i} = \sqrt{\frac{\alpha_{\rm vir} G M_{{\rm cl}, i}}{5 r}},
\end{equation}
where $M_{{\rm cl}, i}$ is user-defined mass of the star cluster from which system $i$ was formed. Beyond this initial sampling, a \codeLink{HydroPopulation}{hydro.pop} has the exact same functionality and methods available as a regular \codeLink{Population}{pop}. A demonstration of evolving a population sampled from a snapshot is given in \tutorialLink{https://cogsworth.readthedocs.io/en/latest/tutorials/hydro/pop.html}{this tutorial}.

\subsection{Visualisation}\label{sec:visuals}

\cogsworth offers several methods for visualising the evolution and end-states of binaries evolved in populations. \cosmic and \gala already provide useful tools for investigating the binary evolution history and galactic orbits, but \cogsworth expands these to aid in the interpretation of simulation data.

\begin{figure}
    \addDocsIcon{fig:cartoon-binary}{https://cogsworth.readthedocs.io/en/latest/auto_examples/plot_cartoon.html}
    \centering
    \includegraphics[width=\columnwidth]{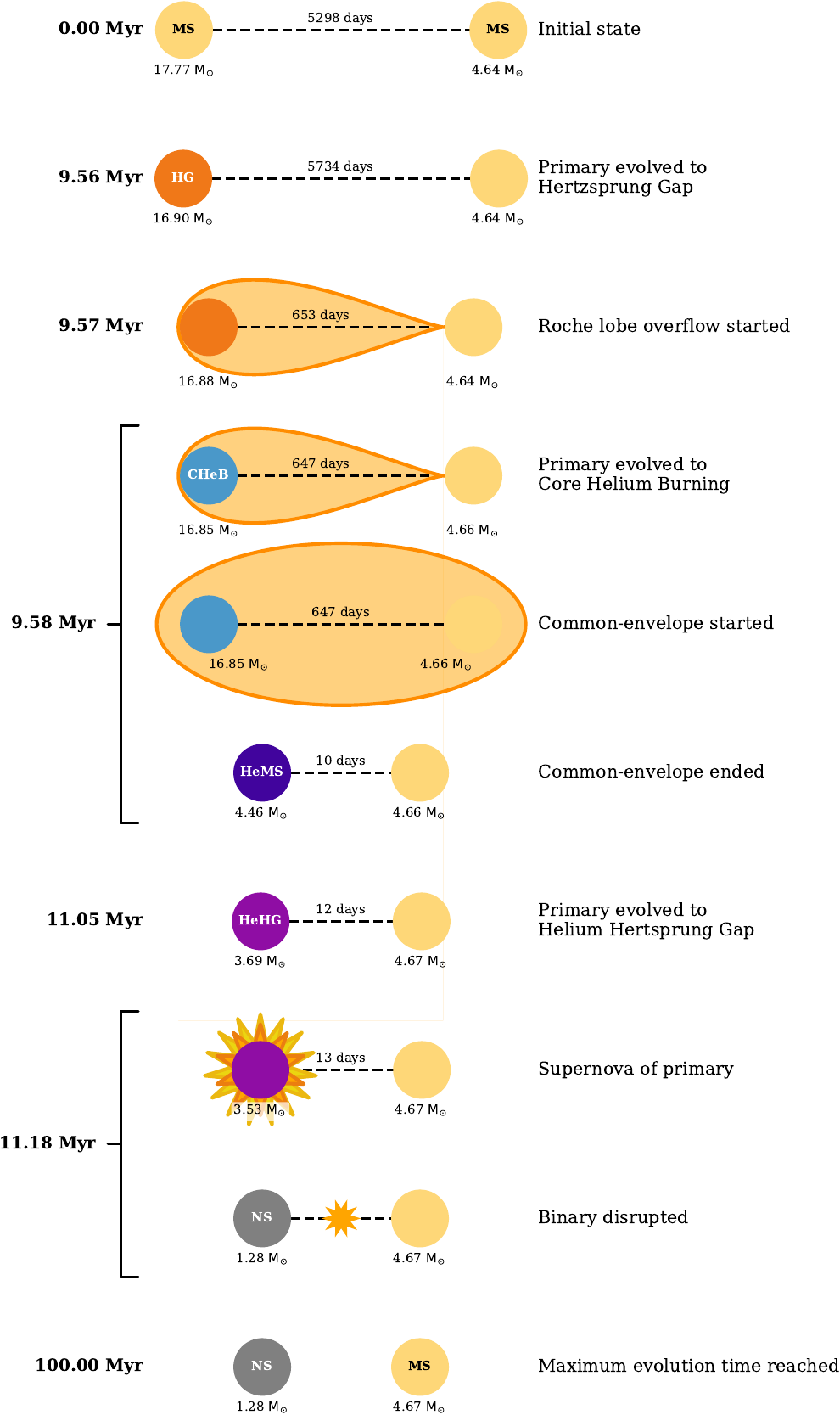}
    \caption{A dynamically generated cartoon binary evolution timeline. Each row shows an evolutionary step, labelled with its time and the event that occurred. Circles are shown for each star, annotated with their masses and the binary's orbital period. A dashed line indicates a bound binary, whilst a lack thereof indicates two unbound stars that were previously in a binary.}
    \label{fig:cartoon-binary}
\end{figure}

\paragraph{Binary evolution} For each binary evolved in \cogsworth, one can dynamically generate a cartoon timeline of its evolution, sometimes called a Van den Heuvel diagram \citep{vandenheuvel:76}, with the function \codestyle{plot\_cartoon\_binary()}. This timeline will show the evolutionary history and is capable of illustrating: the masses of each star, orbital period, mass transfer, common-envelope events, contact phases, mergers and supernovae. The distance between each star in the plot is directly scaled by the orbital separation of the binary (on a log-scale). This functionality provides a simple way to interpret the evolution of a binary without needing to know the meanings of each number representing stellar types and evolutionary stages in \cosmic output tables. We show an example of this in Figure~\ref{fig:cartoon-binary} for a randomly sampled binary that we evolved for 100 Myr. Initially, as the primary loses mass to stellar winds the orbit widens very slightly. However, once the primary ends its main sequence and expands across the Hertzspung gap, it initiates mass transfer. The star continues to evolve during mass transfer, eventually expanding at such a rate to make the mass transfer unstable, leading to a common-envelope and the stripping of the primary star. The primary star reaches supernova after around 11 Myr and the resulting kick disrupts the binary, ejecting the newly formed neutron star and its (prior) companion across the galaxy.

\paragraph{Galactic orbits} \cogsworth provides a wrapper, \codestyle{plot\_orbit()}, on the \gala orbit plotting functionality to allow users to plot projections of a given binary's orbit in galactocentric coordinates. This allows users to plot the orbit of any evolved system, with markers indicating the location of each supernova. For a disrupted system, an additional line will be plotted for the secondary star after the binary disrupts.

\paragraph{Sky maps} For Milky Way simulations, users can plot their simulated populations on the sky with \cogsworth. This is possible either with a simple scatter plot of right ascension and declination, or with a HEALPix Molleweide heatmap via \texttt{healpy} \citep{Zonca2019, 2005ApJ...622..759G}. For HEALPix maps, one can customise the plots in several ways, including choosing the coordinates to plot (celestial, galactic, or equatorial) and the resolution of the map. See Section~\ref{sec:sky_loc_demo} for a demonstration of plotting a simulated population on the sky.

\subsection{Multiprocessing scalability}\label{sec:runtime}

Each binary in \cogsworth is assumed to evolve independently of all other binaries in the simulation (we do not account for dynamical N-body interactions, see Section~\ref{sec:limitations}). An advantage of this is that each system can be efficiently parallelised. Leveraging this fact, \cogsworth uses a multiprocessing pool for the evolution of each binary system. This means that the runtime of simulations scales well with the number of processes used.

\begin{figure}
    \addDocsIcon{fig:runtime-scaling}{https://cogsworth.readthedocs.io/en/latest/tutorials/misc/runtime.html}
    \centering
    \includegraphics[width=\columnwidth]{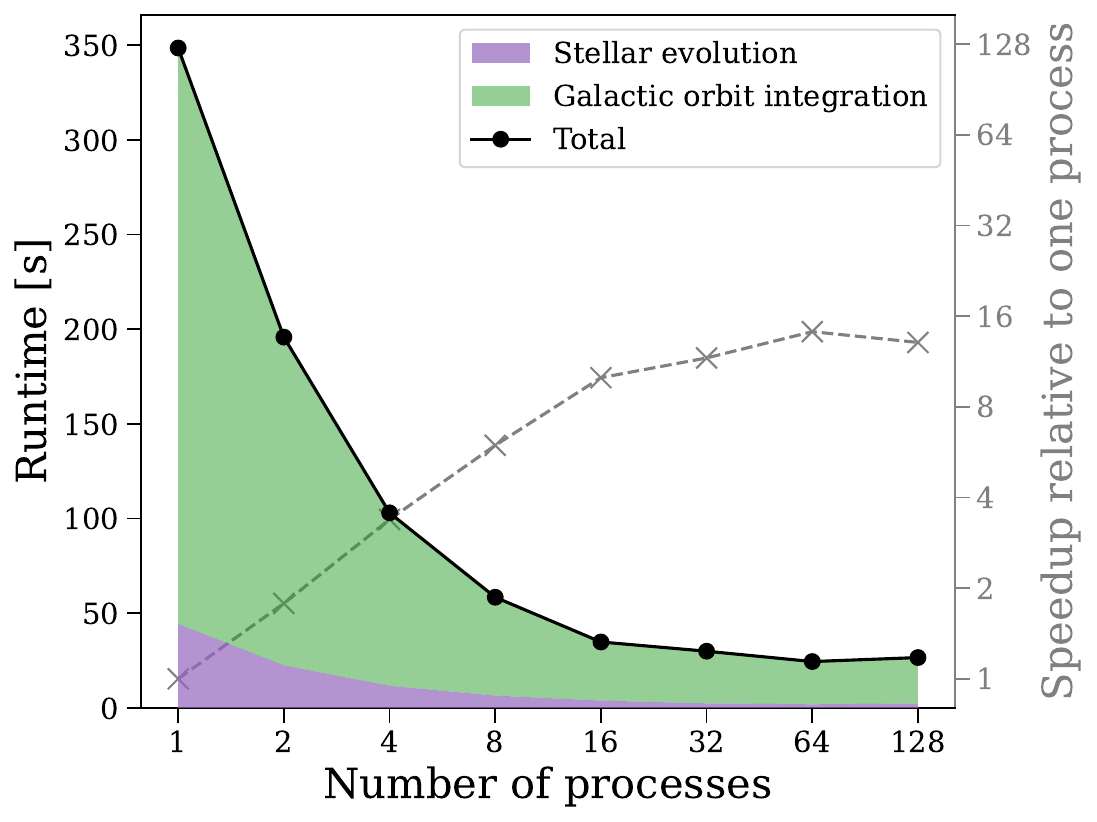}
    \caption{The scaling of \cogsworth's runtime with the number of processes used for a fixed population size of 10,000 binaries. The black points indicate the total runtime, whilst the shaded areas show the relative contribution from the stellar evolution and galactic orbit integration. The right y-axis and grey dashed line shows the relative speedup compared to the simulation using a single process.}
    \label{fig:runtime-scaling}
\end{figure}

We demonstrate the scaling of \cogsworth's runtime with the number of processes used in Figure~\ref{fig:runtime-scaling}. We first sample a fixed population of 10,000 binaries to ensure consistency across runs. We do not include this sampling in the runtime analysis since it is typically negligible relative to the evolution. We computed the runtime for a \cogsworth simulation when using our default settings and varying the number of processes used.

Figure~\ref{fig:runtime-scaling} shows that increasing the number of processes can significantly decrease the runtime of a simulation. However, this only continues up to a point, beyond which increasing the number of processes yields diminishing returns. The reason that adding processes does not always increase the runtime is that small subsets of the population take longer to run, such as binaries with multiple interactions between the two stars, or orbits that pass close to the galactic centre requiring finer time-sampling for the orbital evolution. Since we do not spread these equally among processes, additional resources are left idle while the complicated subset runs on a limited number of cores. In the case of this test, using more than 16 processes does not improve the performance, though we note that this threshold is dependent on the population (both its size and demographics) and the settings chosen by a user. We recommend that users perform similar tests to ascertain the optimal number of processes for their use case.

In Figure~\ref{fig:runtime-scaling} we additionally show the relative contributions to the runtime from performing the stellar evolution and galactic orbit integration. For this example, the galactic orbit integration typically dominates the runtime by around a factor of $\sim$5. The relatively higher cost of the orbit integration is expected given that \cosmic relies on pre-computed stellar models for stellar evolution, whilst \gala fully integrates galactic orbits. The relative contributions from the two phases depend on the simulation that is run. One with more binaries that have many interactions between companions (e.g.\ a population focused on massive stars) may increase the runtime of the stellar evolution. Conversely, a simulation using a more complex galactic potential than the smooth \texttt{MilkyWayPotential2022} (such as one computed from a hydrodynamical zoom-in simulation) may increase the runtime of the galactic orbit integration. We highlight that the overhead added by \cogsworth in connecting these two aspects is negligible.

\subsection{Data storage}\label{sec:data}

\cogsworth includes functions for saving and loading an evolved \codeLink{Population}{pop}. Simulations are efficiently stored in a single HDF5 file using \texttt{h5py} \citep{collette_python_hdf5_2014}, which contains: simulation input settings, a SFH and sampled initial galactic variables, the chosen galactic potential, the initial state, final state and full evolutionary history of the binary stellar population, and the orbits of each system through the galaxy. This provides all the information necessary for  reproducing a simulation and analysing its outputs.

We implement lazy-loading for \cogsworth simulations. This means that not all data is immediately loaded and is instead only loaded as it is needed. For example, one can load a population without the full galactic orbits of the binaries, select a subpopulation of interest (runaway stars, for instance) and plot their orbits, at which point \cogsworth will load only the data necessary for these systems on the fly.

The file size of a simulation depends on various factors. As one would expect, increasing the number of binaries or single stars simulated, increases the size of the output file. Additionally, since the output log includes a row for each significant evolutionary stage, binaries with more interactions between companions result in larger files. The length of time simulated is also an important factor, since it not only allows more time for binaries to experience complex evolution, but also requires more integration timesteps for galactic orbits. Users can reduce the file size of a simulation by specifying that \cogsworth use larger integration timesteps, or even that \cogsworth should only retain the final position of each star rather than its full galactic orbit. As some examples, the population used in Section~\ref{sec:cluster_orbits_use_case} consists of ${\sim}500$ binaries, which is evolved for ${\sim}50$ million years, results in a file size of ${\sim}5$Mb. The population used in Section~\ref{sec:binaries_and_pot_use_case}  is larger, containing 30,000 massive binaries and 30,000 massive single stars, and is evolved for $100\unit{Myr}$, producing a file of ${\sim}300$Mb.

\subsection{Citations}\label{sec:citations}

Given the breadth of work upon which \cogsworth depends, we make it simple for users to ensure they fully credit the work that went into a given simulation. \cogsworth is capable of creating a custom citation statement for any given \codeLink{Population}{pop} using the \href{https://cogsworth.readthedocs.io/en/latest/api/cogsworth.pop.Population.html\#cogsworth.pop.Population.get\_citations}{\codestyle{get\_citations()}} function.

For example, the simplest \cogsworth simulation may cite only \cogsworth itself, \cosmic and \gala. But if you compute the observable features of your population, \cogsworth will add citations for the dust maps, isochrones, and selection function that you use. This is similarly true for citations regarding star formation histories and hydrodynamical zoom-in simulations.

\newpage

\section{Use cases}\label{sec:use_cases}

The following subsections each demonstrate a particular use case of \cogsworth, showcasing its capabilities in binary stellar evolution, galactic dynamics, observables estimation, and integration with hydrodynamical simulations. Each \tutorialIcon{} icon links directly to a page in our online documentation that guides users through using \cogsworth to reproduce a given figure.

Unless otherwise specified in the individual use cases, each \cogsworth simulation uses the \codeLink{Wagg2022}{sfh} SFH, the \texttt{MilkyWayPotential2022} Galactic potential and the default binary physics settings from \cosmic v3.4.16. Additionally, primary masses are sampled using the \citet{Kroupa+2001:2001MNRAS.322..231K} IMF, a uniform mass ratio distribution is assumed and the initial orbital period and eccentricity distributions follow \citet{Sana+2012}. These assumptions are appropriate for massive stars, which can experience core-collapse events and be kicked and disrupted.

\subsection{The importance of binary evolution and the galactic potential}\label{sec:binaries_and_pot_use_case}

In this use case, we demonstrate the need for accounting for binary interactions and a galactic potential simultaneously. We use \cogsworth to simulate a population of massive stars formed in the most recent 1 Gyr in the Milky Way assuming a 50\% binary fraction. We then repeat the orbital integration for the binaries in this population, but without a galactic potential. In this way, binaries with no core-collapse events remain in their birth locations, whilst those that receive kicks continue at their ejection velocity indefinitely.

\begin{figure}
    \addDocsIcon{fig:zgrid}{https://cogsworth.readthedocs.io/en/latest/case_studies/binaries_and_potentials.html}
    \centering
    \includegraphics[width=\columnwidth]{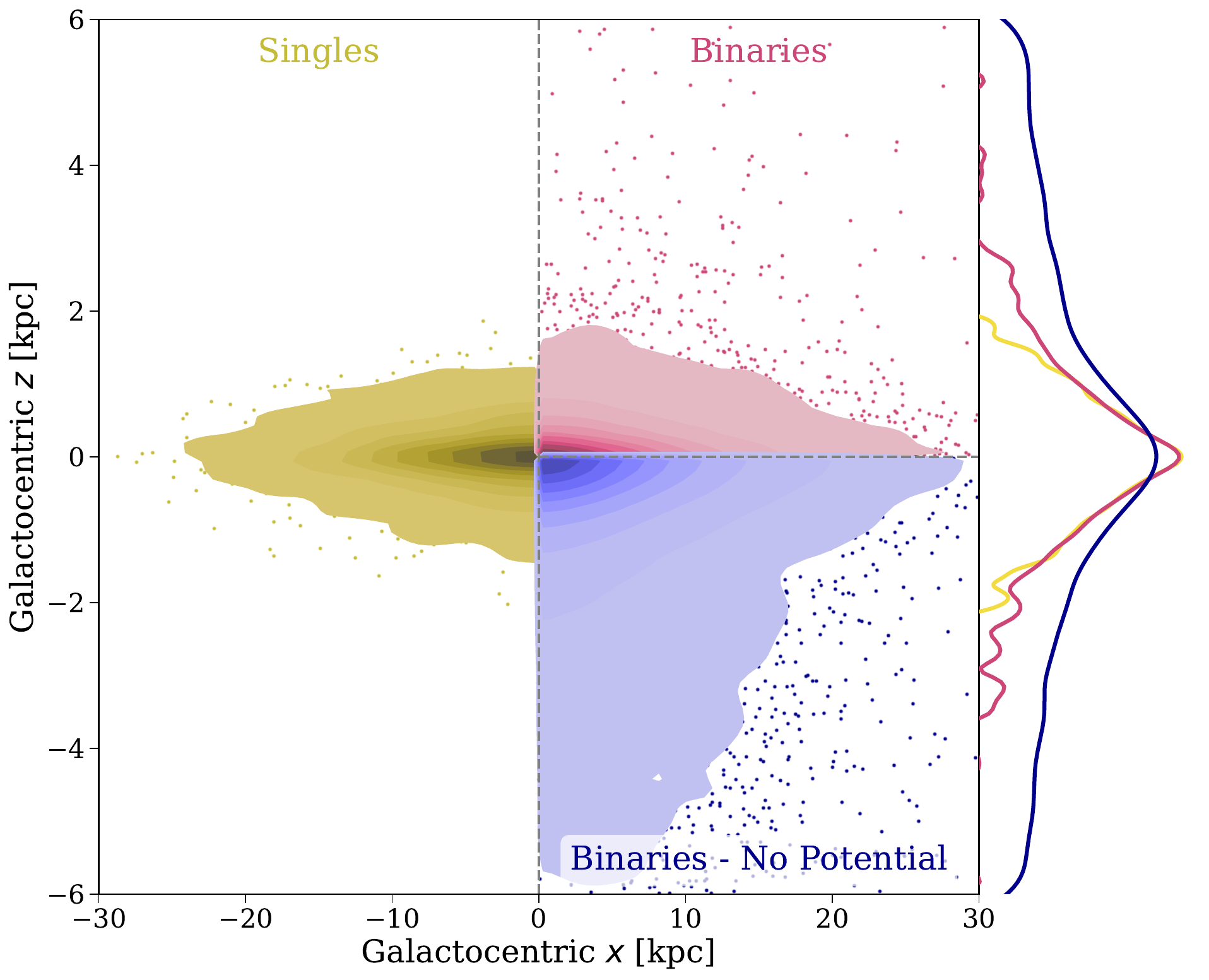}
    \caption{Binary evolution and the presence of a galactic potential both have a significant impact on the spatial distribution of stars. Each panel shows the spatial distributions of massive stars at present day from a \cogsworth population of single stars (left), binaries (right, top) and binaries evolved without a galactic potential (right, bottom). Density distributions are computed using kernel density estimators (KDE) and shown (in linearly spaced isolevels) up to the 98th percentiles, with the remaining stars plotted as scatter points. Marginal KDEs are shown on the right with a log scale.}
    \label{fig:zgrid}
\end{figure}

We compare the present-day positions of single stars and binary stars in Figure~\ref{fig:zgrid}, additionally showing the distribution of binary stars when neglecting the Galactic potential. First, comparing the single and binary stars in the presence of a potential, we note that the tails of the binary distribution are significantly extended. For single stars, the fraction of the population at $|z| > 1 \unit{kpc}$ is only 0.9\%, whereas for binaries it increases by a factor of 2x to 1.9\%. Previous work has investigated the spatial distribution of compact objects without accounting for binary interactions \invisibleedit{, assuming that these interactions do not change the overall distributions \citep{underworld}. However, as Figure~\ref{fig:zgrid} shows, our \cogsworth simulations demonstrate that binary interactions can significantly alter the spatial distributions of massive stars and compact objects.}

\begin{figure*}
    \addDocsIcon{fig:pot-sigma-vars}{https://cogsworth.readthedocs.io/en/latest/case_studies/sigma_plus_potentials.html}
    \centering
    \includegraphics[width=\textwidth]{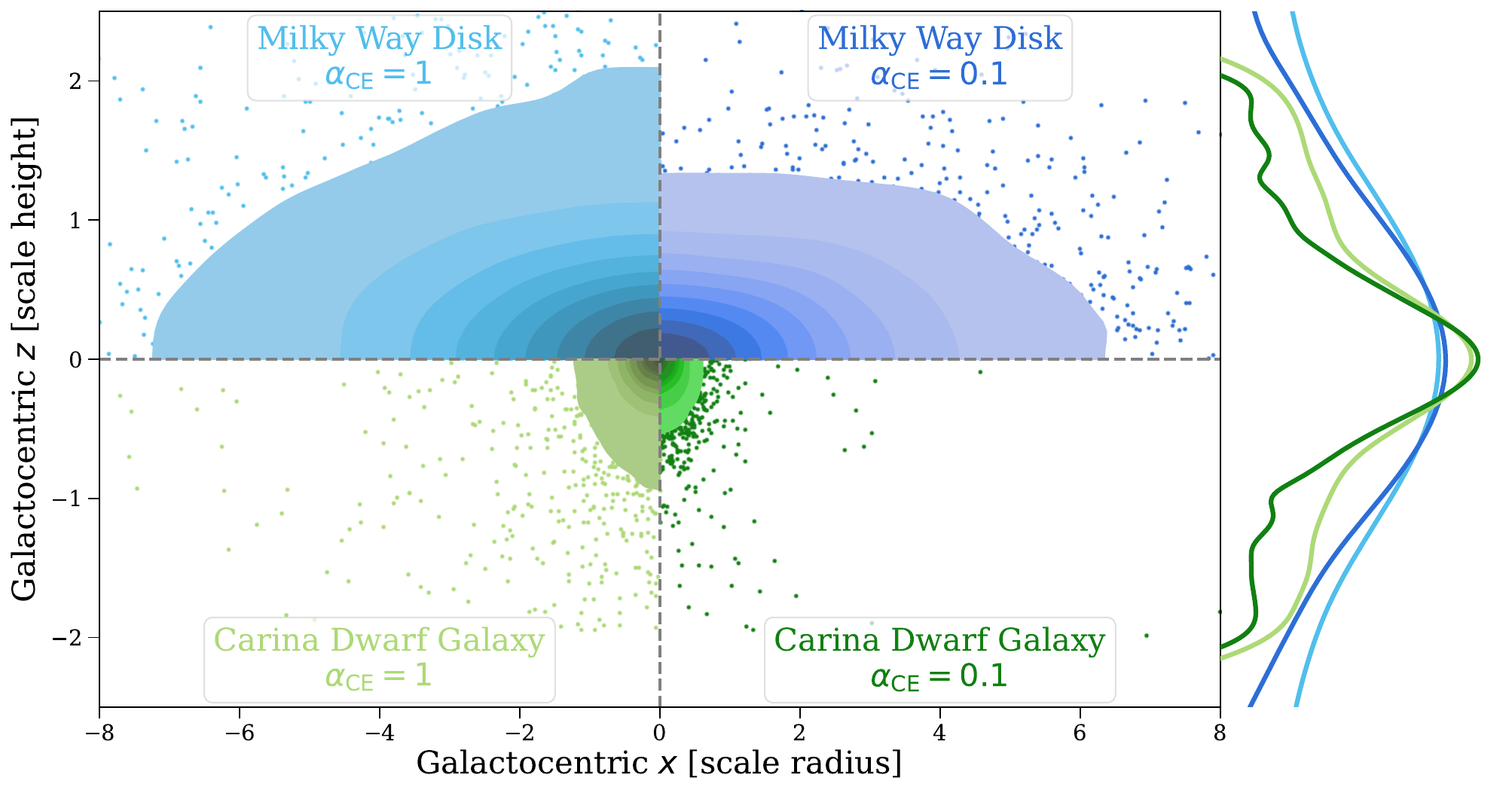}
    \caption{Similar to Figure~\ref{fig:zgrid}, but comparing the impact of varying binary physics and the galactic potential. The axes are now in units of scale radius and scale height for better comparison between the two galaxies. Each quadrant corresponds to a different combination of common-envelope efficiency and galactic potential (annotated in each).}
    \label{fig:pot-sigma-vars}
\end{figure*}

Moreover, comparing the binary population with and without the Galactic potential, it is clear that neglecting the Galactic potential results in misleading conclusions regarding spatial distributions. In particular, the population without a potential is broader in both width and height, and the fraction at $|z| > 1 \unit{kpc}$ is $8.6\%$, more than 4x the fraction when accounting for the potential.

\subsection{Comparing the impact of supernova kicks and galactic potentials on spatial distributions}

A key feature of \cogsworth is its ability to self-consistently account for the effect of binary interactions and galactic potentials. We explore this capability by determining the relative impact of varying supernova natal kicks and galactic potentials on the spatial distribution of a population of binary stars.

\invisibleedit{We vary the efficiency of common-envelope phases from our default choice of $\alpha_{\rm CE} = 1.0$ to an extreme choice of $\alpha_{\rm CE} = 0.1$, where $\alpha_{\rm CE}$ is the fraction of orbital energy that is available to unbind the envelope \citep{Webbink+1984:1984ApJ...277..355W, deKool+1990:1990ApJ...358..189D}. This means that when $\alpha_{\rm CE}$ is lower equivalent systems that survive the phase are tighter after a common-envelope, such that they have higher orbital velocities and therefore ejection velocities. However, binaries are also more likely to merge as a result of a common-envelope phase.} For galactic potentials, we compare the Milky Way's disk (using parameters from \citealt{Sanders+2015:2015MNRAS.449.3479S}) to a spheroidal dwarf galaxy with parameters matching the Carina dwarf galaxy \citep{Pascale+2019}. The latter is assumed to be in isolation, such that there is no tidal stripping from the Milky Way as is observed in the Carina galaxy.


We sample an initial binary population (assuming a binary fraction of 100\%) from the most recent 1 Gyr of star formation, retaining only systems with primary stars more massive than $7 \unit{M_\odot}$, since here we are only interested in binaries that experience a core-collapse event. \cosmic reports the precise initial conditions of sampled populations, making them easily reproducible with different evolution settings. Using this feature, we evolve identical initial populations with the two different \invisibleedit{common-envelope efficiencies}, then integrate the orbits of each variation through the two different potentials. This results in 4 different populations of present-day positions.

In Figure~\ref{fig:pot-sigma-vars}, we compare these populations, showing the positions of all massive stars that experienced a core-collapse event, or were a companion to a star that did. \invisibleedit{To make the comparison across galaxies easier, we plot the positions in terms of the scale radius, $r_s$, and scale height, $z_s$, of each galaxy, where for the Milky Way we assume $r_s = 2.6 \unit{kpc}$ and $z_s = 0.3 \unit{kpc}$ \citep{Bland-Hawthorn+2016:2016ARA&A..54..529B} and for the Carina galaxy we assume $r_s = z_s = 1 \unit{kpc}$ \citep{Pascale+2019}.} Given the variety of results from the four panels, we highlight that both binary physics and the choice of galactic potential can have a strong effect on the resulting spatial distribution of massive stars. \invisibleedit{In particular, less efficient common-envelope phases result in reduced galactic scale heights and radii for massive stars. This is because a significant fraction of binaries that experience a common-envelope merge during the event, thus reducing the number of stars that are ejected from their binaries. Although the width of the overall distributions are difficult to statistically distinguish, the tails of the distribution are more strongly affected. The fraction of objects at $|z| > z_s$ decreases from 5.1\% to 3.0\% for the disc population, and from 1.4\% to 0.4\% for the dwarf galaxy population.} Therefore, observing the outliers in a galactic height distribution of massive stars can be more informative than the overall width of the distribution for inferring the efficiency of common-envelopes. One would need to also investigate other parameters that could affect these distributions, in particular the critical mass ratio for the stability of mass transfer and the efficiency of stable mass transfer \citep[e.g.,][]{Evans+2020:2020MNRAS.497.5344E}.

\newpage

\subsection{Evolution of binary orbits in a star cluster}\label{sec:cluster_orbits_use_case}

\cogsworth is capable of producing populations of binaries based on hydrodynamical simulations (Section~\ref{sec:hydro_linking}). We use \cogsworth to post-process the FIRE m11h simulation \citep{El-Badry+2018:2018MNRAS.473.1930E_m11h, Wetzel+2023:2023ApJS..265...44W}, an intermediate-mass halo with a strong disc component, fitting a galactic potential and rewinding all star particles formed in the past $150 \unit{Myr}$ to their formation locations. We then sample a binary population from each star particle, matching its formation time, metallicity and mass. For this example, we examine one random star particle and the orbits of its constituents. This star particle was formed ${\sim}43\unit{Myr}$ before present-day, with a mass of ${\sim}6000\unit{M_\odot}$ and a metallicity of $Z \approx 0.0137$. The sampled population consists of ${\sim 7600}$ systems, split evenly between single stars and binary stars (as one would expect given our assumption of a 50\% binary fraction). Note that we neglect the self-gravity of the cluster.

\begin{figure}
    \addDocsIcon{fig:orbits-example}{https://cogsworth.readthedocs.io/en/latest/case_studies/fire.html}
    \centering
    \includegraphics[width=\columnwidth]{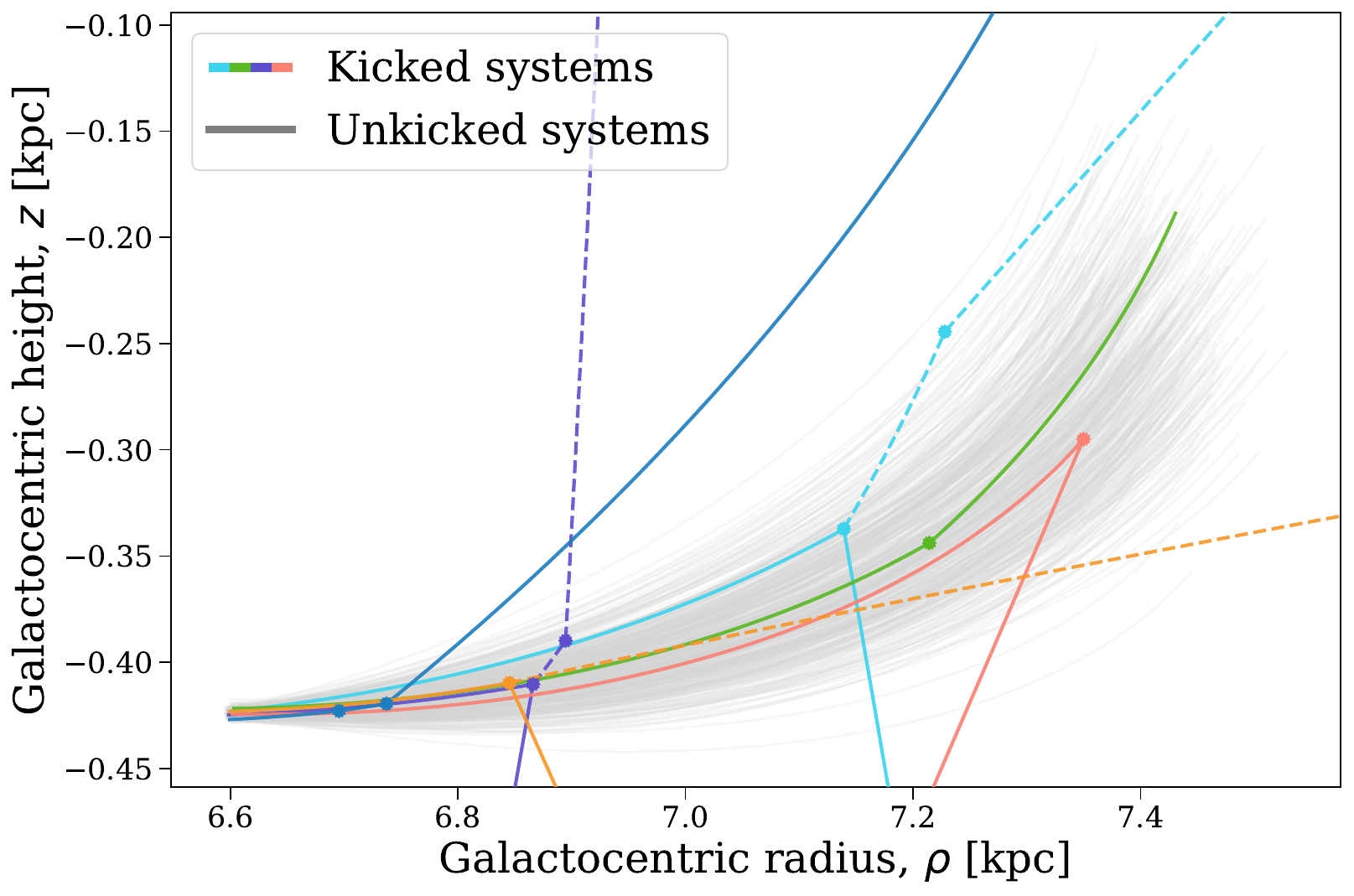}
    \caption{An example star cluster, sampled from a single star particle in the FIRE m11h hydrodynamical zoom-in simulation, evolved within a \cogsworth simulation. Each line shows the orbit of a star in the cluster. Coloured lines are for stars that had supernova events (highlighted by scatter points), while the rest are shown in grey. For binaries that disrupt, an additional dashed line is shown for the subsequent orbit of the ejected companion.}
    \label{fig:orbits-example}
\end{figure}

In Figure~\ref{fig:orbits-example}, we plot the orbits of a representative subset of 500 of the binaries sampled from the star particle. In grey, we show the orbits of binaries that experienced no supernovae events, which are by far the majority since the IMF favours low mass stars. The cluster is formed in the lower left (at $\rho = 6.6 \unit{kpc}, z = -0.42 \unit{kpc}$) and evolves to larger $\rho$. One can note the dissolution of the cluster over time in the grey orbits, which occurs as a result of the initial velocity dispersion (Eq.~\ref{eq:vel_disp_cluster}).

The coloured lines show the more eventful orbits of binaries that experienced supernovae. In many cases this leads to the disruption of the binary orbit and so we show the orbit of the subsequent evolution of the ejected companion with a dashed line. With \cogsworth, one can examine the detailed evolution of each binary to understand its orbit. The examples shown include several scenarios involving bound, disrupted and merged binaries - we discuss each in detail below.

The earliest core-collapse event occurs for the dark blue binary after $4 \unit{Myr}$, which is indicated by closest scatter point to the cluster origin in the lower left. This forms a $12 \unit{M_\odot}$ black hole and, due to a fallback fraction of $92\%$ for the black hole, much of the explosion asymmetry is negated, resulting in a relatively weak natal kick of $11 \unit{km}{s^{-1}}$, which allows the binary the remain bound. The companion to this binary reaches core-collapse $1.5 \unit{Myr}$ later, forming a slightly less massive black hole of $6 \unit{M_\odot}$, with a stronger natal kick of $70 \unit{km}{s^{-1}}$. Yet the binary is much tighter at this point (with a separation of $45 \unit{R_\odot}$) and thus has a higher binding energy. This means that it remains bound and is ejected from the cluster as a binary black hole.

The first supernova in the orange binary occurs $9\unit{Myr}$ after the cluster birth and forms a neutron star with a natal kick of $447 \unit{km}{s^{-1}}$. This kick disrupts the binary orbit, such that both stars are ejected from the cluster. The secondary is a lower mass star of $3.6\unit{M_\odot}$ and so experiences no supernova, but it is ejected from the cluster at $160 \unit{km}{s^{-1}}$ and as such is now a runaway star.

The purple binary experiences its first supernova at $9.3 \unit{Myr}$ and, similar to the orange binary, this forms a $1.6 \unit{M_\odot}$ NS with a natal kick of $415 \unit{km}{s^{-1}}$ that unbinds the binary. Interestingly, the mass ratio of this system is inverted, such that the companion forms a more massive $5.3 \unit{M_\odot}$ black hole after its core-collapse $2 \unit{Myr}$ later. This inversion occurred as a result of significant, near conservative mass transfer from the primary star during its Hertzsprung gap phase $1.2 \unit{Myr}$ before it reached core-collapse. Both supernovae for the light blue binary form neutron stars (of $1.3$ and $2.2 \unit{M_\odot}$ respectively) with strong kicks (of $406$ and $728 \unit{km}{s^{-1}}$ respectively), which disrupt the binary and eject both neutron stars rapidly from the cluster.

\begin{figure*}
    \addDocsIcon{fig:ZRt}{https://cogsworth.readthedocs.io/en/latest/auto_examples/plot_ZRt_Wagg2022.html}
    \centering
    \includegraphics[width=0.49\textwidth]{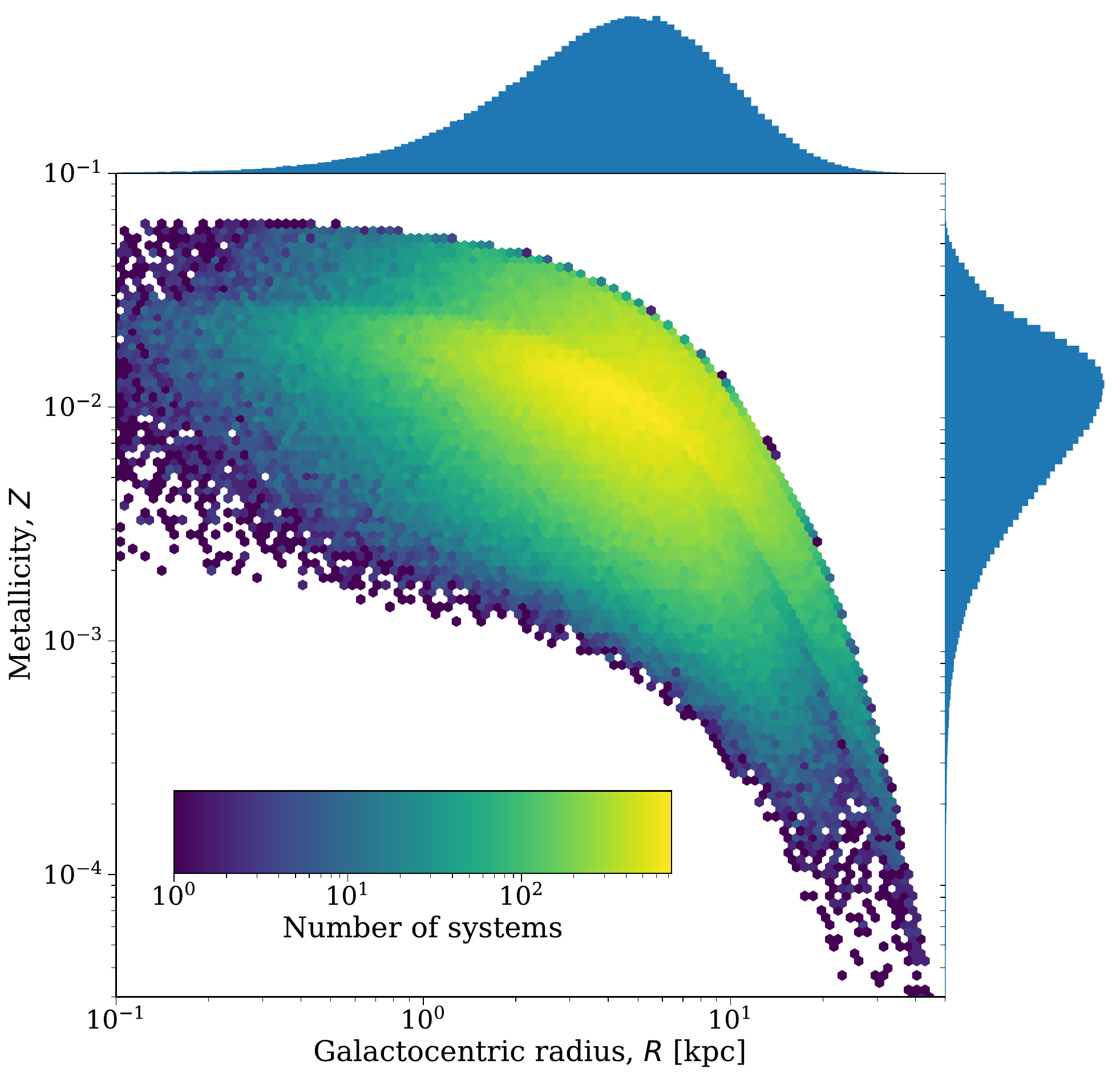}
    \includegraphics[width=0.49\textwidth]{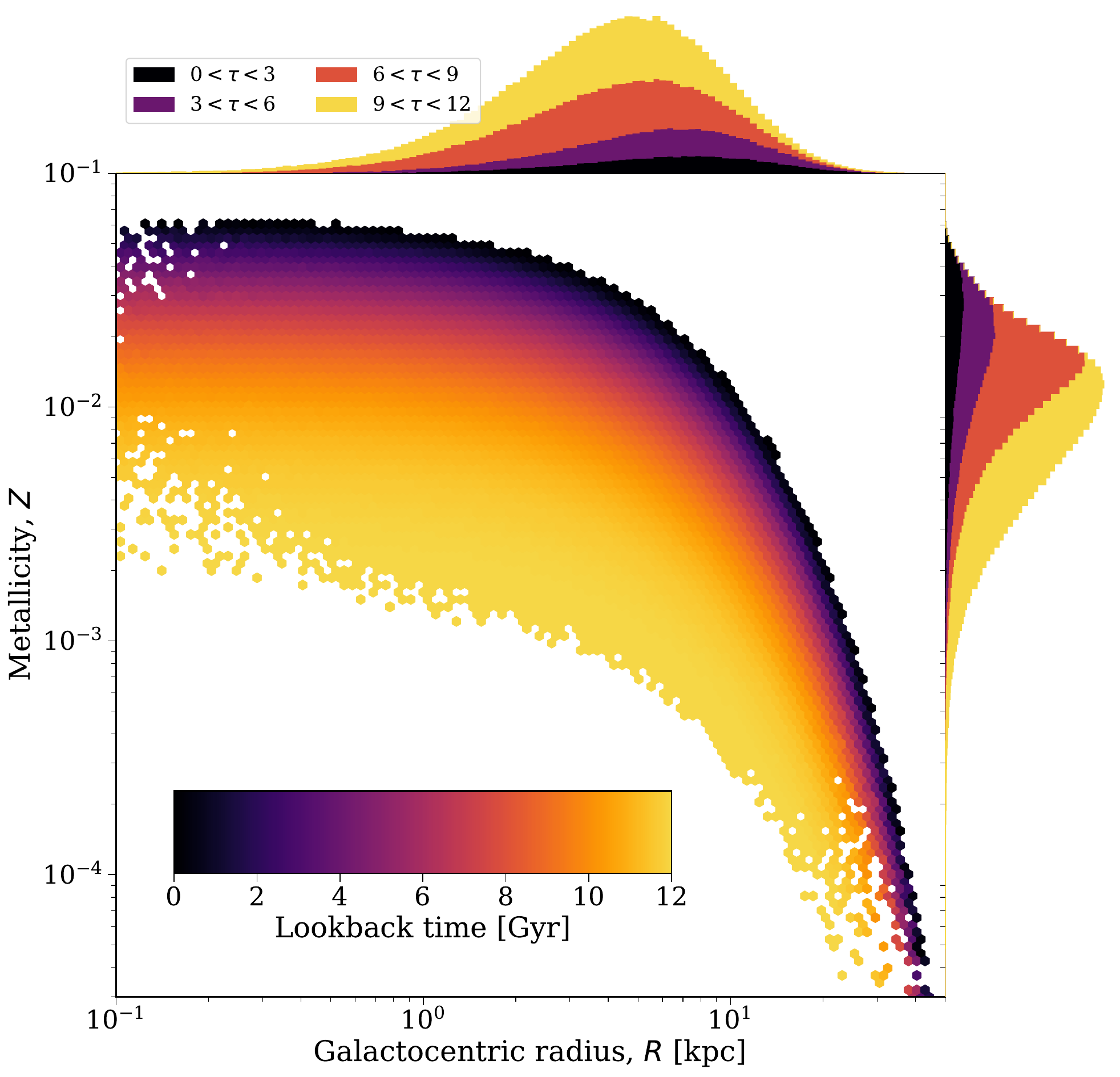}
    \caption{The relationship between galactocentric radius, metallicity and lookback time in the \codeLink{Wagg2022}{sfh} SFH. \textbf{Left:} The main panel shows a 2D histogram of the number of stars formed at a given radius and metallicity, with marginal distributions for each parameter shown as histograms on the side panels. \textbf{Right:} The main panel shows a 2D histogram in which each bin is coloured by the average lookback time of associated stars. Marginal distributions are now shown as stacked histograms, grouped by lookback time.}
    \label{fig:ZRt}
\end{figure*}

The primary star in the green binary reaches core-collapse $26 \unit{Myr}$ after the cluster birth, forming a neutron star of $1.27 \unit{M_\odot}$. The star explodes as an electron-capture supernova and thus its kick is assumed to be weaker, in this case the drawn natal kick is only $35 \unit{km}{s^{-1}}$ and the binary remains bound. In addition, as a result of the angle of the kick relative to the binary's orbit, this only results in a $3.4 \unit{km}{s^{-1}}$ change to the systemic velocity of the binary. As a result, the binary remains bound to the cluster for its subsequent evolution.

Finally, after ${\sim}33 \unit{Myr}$, the primary star in the red binary finishes its main sequence, As it expands during its Hertzsprung gap phase, it overflows its Roche lobe, causing unstable mass transfer which leads to a merger. The merged star then reaches supernova $4 \unit{Myr}$ later, forming a neutron star which is ejected by its strong natal kick of $819 \unit{km}{s^{-1}}$, in almost the opposite direction to the cluster's centre of mass motion.

\subsection{Examining metallicity-radius-time relations in star formation histories}\label{sec:sfh_use_case}

\cogsworth can be used to sample detailed star formation histories independently of evolving binary stars or performing galactic orbit integration (see Section~\ref{sec:galactic_SFH}). In this use case, we explore the \codeLink{Wagg2022}{sfh} SFH model in more detail. We sample 500,000 points (which could be designated as a single or binary star) from the SFH, which each have an associated position, lookback time and metallicity.

In the left panel of Figure~\ref{fig:ZRt}, we plot the distribution of Galactocentric radii and metallicities for each sampled point. As a general trend, one can note that stars closer to the centre of the Galaxy are more metal-rich than those on the outer edges. This is due to the inside-out growth of the Galaxy \citep[e.g.,][]{Fall+1980:1980MNRAS.193..189F, Frankel+2019:2019ApJ...884...99F}, which is accounted for in this SFH following the model of \citet{Frankel+2018}. Additionally, the discontinuity in the distribution (occurring at inner radii at $Z \approx 0.03$) is a result of the multi-component nature of the model. The upper right portion above the discontinuity comes is primarily from the low-$[\alpha/{\rm Fe}]$ disc component, which forms stars from $8 \unit{Gyr}$ ago until present-day, while the lower portion is primarily from the high-$[\alpha/{\rm Fe}]$ disc, which form stars from $12 \unit{Gyr}$ ago until $8 \unit{Gyr}$ ago. The bulge component contributes to both portions, though only at small radii.

For a given radius, there is a wide variation in the metallicity of sampled stars. This is because of the birth time of each star, which we demonstrate in the right panel of Figure~\ref{fig:ZRt}. The 2D histogram now shows the average lookback time, $\tau$, of the stars in each bin (where $\tau = 0$ corresponds to present-day). The clear gradient shows that over time the Galaxy as a whole becomes more metal-rich as it is enriched by stellar evolution. This is additionally visible in the marginal distribution of metallicities, where no high metallicity stars are formed at early birth times. The marginal distribution of radii again demonstrates the inside-out growth, as older stars were formed closer to the Galactic centre.

\subsection{Simulating a \gaia colour-magnitude diagram}

In this use case, we highlight \cogsworth's ability to transform an intrinsic population into a simulated observable population (see Section~\ref{sec:observables}). We sample 2500 binary systems over the full SFH of the Milky Way and evolve them until present day.

We compute observables for this population using the \href{https://cogsworth.readthedocs.io/en/latest/api/cogsworth.pop.Population.html#cogsworth.pop.Population.get_observables}{\codestyle{Population.get\_observables()}} function in \cogsworth. We first calculate the absolute magnitude of each star and determine which star is brighter in a binary. \cogsworth then converts these magnitudes to the \gaia filters G, BP and RP using the MIST isochrones to apply bolometric correction with the \texttt{isochrones} package \citep{Morton+2015:2015ascl.soft03010M, Dotter+2016, Choi+2016:2016ApJ...823..102C, Paxton2011, Paxton2013, Paxton2015}. Finally, \cogsworth uses the \texttt{dustmaps} package to account for dust extinction through the application of the Bayestar 2019 dust maps \citep{2018JOSS....3..695M, Bayestar}.

\begin{figure}
    \addDocsIcon{fig:gaia-cmd}{https://cogsworth.readthedocs.io/en/latest/auto_examples/plot_cmd.html}
    \centering
    \includegraphics[width=\columnwidth]{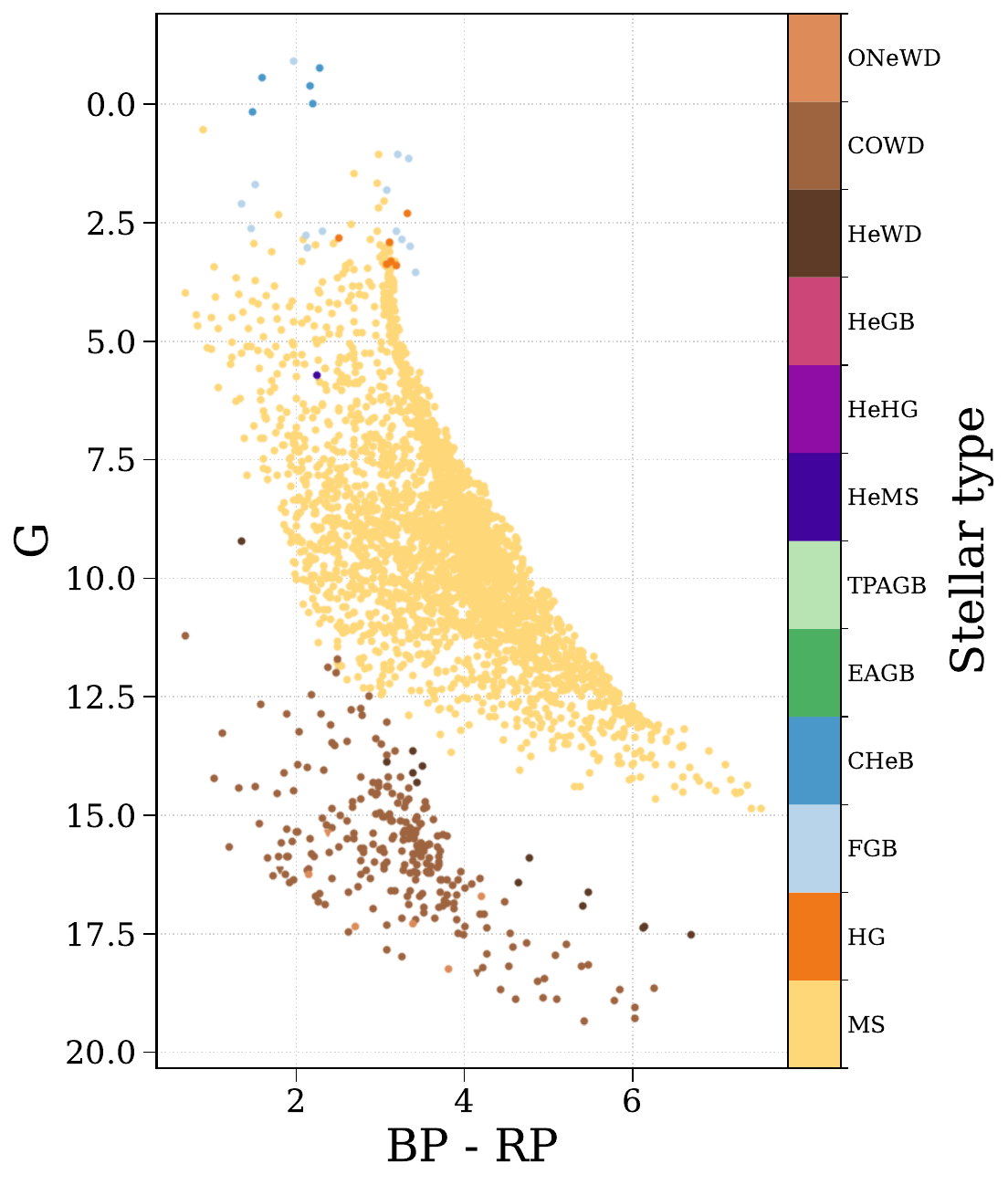}
    \caption{An example simulated \gaia colour-magnitude diagram. Each scatter point corresponds to a binary or disrupted star, coloured by its stellar type (using the stellar type of the brighter component for binaries). The stellar types abbreviated in the colourbar follow those defined in \texttt{BSE} \citep[see Section 4 of][]{Hurley+2000:2000MNRAS.315..543H}.}
    \label{fig:gaia-cmd}
\end{figure}

We plot the resulting colour-magnitude diagram (CMD) in Figure~\ref{fig:gaia-cmd}, colouring each bound system by the stellar type of the brighter component in the $G$ band. These systems cover a range of metallicities, distances and ages and hence have a wide spread in the CMD. In addition to these systems, 59 isolated (either from mergers or binary disruptions) neutron stars and black holes are present in the evolved population. This same simulation could be easily repeated with a starburst localised in one specific place to model a specific cluster CMD instead.

\subsection{Comparing present-day sky locations with and without supernova kicks}\label{sec:sky_loc_demo}

\cogsworth can report the present-day sky position of each source, in addition to its evolutionary history. In this use case we demonstrate how one can track the relative positions of binary companions after they disrupt, as well as consider where they would be found if no supernova kick had occurred.

We sample and evolve 100 random binaries in the Milky Way with our default assumptions, except we set the minimum mass of the IMF to $3 \unit{M_\odot}$. This preferentially samples more massive stars, which are more likely to reach core-collapse and cause a binary disruption. From this population, we subselect five binaries that are disrupted. These binaries experienced at least one core-collapse event which disrupted the orbit and led to the separation of the two unbound companions. We use \cogsworth to compute the present-day sky location of both companions for each binary, before re-integrating the binary's orbit without accounting for the impact of supernova kicks. \cogsworth returns the final coordinates of stars as an \texttt{Astropy SkyCoord} \citep{astropy:2013, astropy:2018, astropy:2022}, which allows for simple transformation between coordinate frames.

\begin{figure}
    \addDocsIcon{fig:disrupted_pairs}{https://cogsworth.readthedocs.io/en/latest/auto_examples/plot_kick_vs_nokick_lb.html}
    \centering
    \includegraphics[width=\columnwidth]{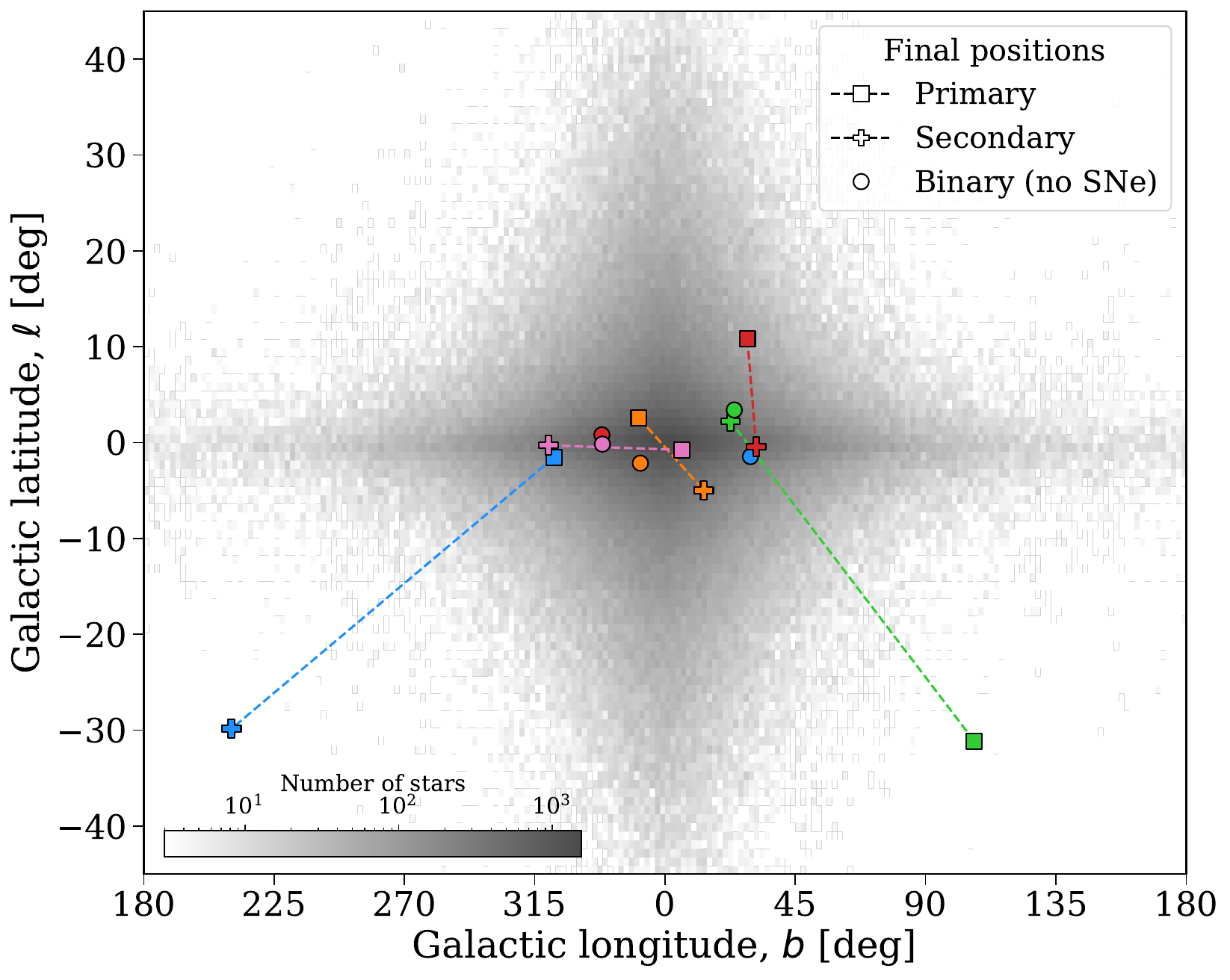}
    \caption{Present-day sky positions of disrupted binaries, with and without supernova kicks. A square (plus) marker shows the location of the primary (secondary) star from a disrupted binary, connected to its companion by a dashed line. Circular markers indicate the location of the binary had no supernova kicks occurred. The background density shows a 2D histogram of 500,000 stars sampled from the same SFH (\codeLink{Wagg2022}{sfh}) for comparison. Note that disk appears extended in latitude because we limit the axes to the relevant region.}
    \label{fig:disrupted_pairs}
\end{figure}

In Figure~\ref{fig:disrupted_pairs}, we compare the present-day sky locations (in galactic coordinates) of each compact object from the disrupted binaries to the location of the binary had it experienced no supernova kicks. As expected, when neglecting supernova kicks, binaries are generally concentrated close to the galactic midplane and centre. However, when accounting for the effect of supernova kicks on the internal and galactic orbits, the present-day sky locations are often significantly different. In many cases, companions are located not only far from one another, but also far from the position of the binary had no kicks occurred. Several of the compact objects from the disrupted binaries are found well beyond the typical sky locations of galactic sources (shown as a histogram in the background), though all remain bound to the galaxy. In particular, the secondary from the blue binary is first ejected from the binary as a runaway star (travelling at ${\sim}35 \unit{km}{s^{-1}}$) after the primary reaches core-collapse. After spending 10 Myr as an O-type runaway star, the secondary reaches core-collapse and is kicked onto an orbit that takes it even further from the typical galactic population. The green binary starts on a relatively wide initial orbital period of ${\sim}30,000 \unit{days}$. The binary widens by nearly $10\%$ due to the stellar winds of the primary before this star's core-collapse $32 \unit{Myr}$ after the birth of the binary. The primary forms a neutron star that receives a natal kick of $325 \unit{km}{s^{-1}}$, such that it takes large excursions from the galactic plane during its orbit. However, due to the large orbital period at the supernova, the secondary star is only ejected at $16 \unit{km}{s^{-1}}$. This star eventually forms a white dwarf and receives no further kick, hence its present-day location is relatively similar to that of the binary had no kick occurred.





\section{Limitations}\label{sec:limitations}

Although \cogsworth has a wide range of features and capabilities, there are still some limitations to the code that users should be aware of.

\paragraph{Dynamical interactions} We do not implement dynamical interactions between systems or account for any N-body dynamics. As mentioned in Section~\ref{sec:runtime}, each binary in \cogsworth is evolved independent of all others. This means that dynamical formation channels for different populations (such as interactions in dense clusters forming gravitational-wave progenitors) cannot be simulated in \cogsworth. However, we do intend to implement a prescription for emulating dynamical cluster ejections (see Section~\ref{sec:cluster_ejections}), such that investigations into runaway stars could consider both channels.

\paragraph{Impact of galactic potential on internal orbits} \cogsworth accounts for the impact of stellar evolution on the galactic orbits of binaries (i.e.\ as a result of supernovae). Yet it does not currently account for the inverse case, in which a galactic potential alters the orbit of the binary. For wide binaries the potential can significantly affect the orbit, causing large-amplitude oscillations and potentially drive systems to disrupt or merge \citep[e.g.,][]{Weinberg+1987:1987ApJ...312..367W,Heisler+1986:1986Icar...65...13H,Jiang+2010:2010MNRAS.401..977J,Modak+2023:2023MNRAS.524.3102M,Stegmann+2024:2024arXiv240502912S}. However, for closer binaries the effect is negligible and as such as we do not currently account for it in \cogsworth.

\paragraph{Population synthesis model uncertainties} \cogsworth uses \cosmic for binary population synthesis, which is a code based on \texttt{BSE}. The \texttt{BSE} code relies on approximate parametric prescriptions for a limited set of evolutionary tracks of single stars \citep{pols:98, Hurley+2000:2000MNRAS.315..543H,Hurley+2002}. Although many of the original prescriptions used in the \texttt{BSE} code have been improved in \cosmic \citep{COSMIC}, the core of the code still relies on the same methodology. In particular, the treatment of mass loss and the stability of mass transfer, as well as the reliability of the most massive progenitor models, is uncertain.
However, some of these uncertainties can be alleviated by incorporating information on the internal structure of stars \citep[e.g.,][]{Kruckow+2018:2018MNRAS.481.1908K, Fragos+2023:2023ApJS..264...45F}. \cosmic is in the process of being integrated with \texttt{METISSE}, MEthod of Interpolation for Single-Star Evolution. \texttt{METISSE} is an alternative to fitting formulae that allows for the interpolation between pre-computed detailed one-dimensional stellar evolution tracks, while maintaining the same code interfaces as the previously implemented prescriptions of SSE \citep{Agrawal+2020:2020MNRAS.497.4549A, Agrawal+2023}.
By working with updated libraries of pre-computed single star tracks from \texttt{MESA} \citep{Paxton2011, Paxton2013, Paxton2015, Paxton2018, Paxton2019, Jermyn2023}, \texttt{METISSE} enables a wide range of investigations of the impact of uncertainties in single-star evolution like convection, rotation, and nuclear reaction rates and how these uncertainties interface with uncertainties in binary interaction physics. Once \texttt{METISSE} is fully integrated into \cosmic, \cogsworth will be able to immediately leverage these new improvements.

\section{Future developments}\label{sec:future}

We intend to complete further development on \cogsworth beyond this initial release. In the following subsections we highlight some areas in which we aim to focus.

\subsection{Time-evolving galactic potentials}\label{sec:time-evolving-potentials}
Traditional models using static galactic potentials are not capable of describing the dynamically complex evolutionary history of galaxies, and can lead to misleading results \citep[e.g.,][]{Arora+2022}. Although this is less relevant for shorter lived populations (such as massive runaway stars), it could have important implications for longer lasting tracers of binary endpoints (e.g., gravitational wave mergers). Thus we also intend to leverage our integration with hydrodynamical simulations to implement a \cogsworth option for a time-evolving gravitational potential that accounts for the mass growth of a galaxy over time.

\subsection{Dynamical cluster ejections}\label{sec:cluster_ejections}

\cogsworth does not currently account for the impact of dynamical interactions between binaries in dense environments, as noted in Section~\ref{sec:limitations}. The interactions can change the initial architecture of binaries \citep[e.g.,][]{Fujii+2011:2011Sci...334.1380F} and create alternate formation channels for binary products. For instance, runaway stars are thought to be formed in two main channels: the disruption of binaries as a result of supernovae \citep{Blaauw+1961, Eldridge+2011:2011MNRAS.414.3501E, Renzo+2019:2019A&A...624A..66R} and dynamical ejections from stellar clusters \citep{Poveda+1967}. Fully modelling the latter channel would require more complex N-body dynamics that are currently beyond the scope of the code. Instead, we intend to create an approximation in which we will give a fraction of massive stars kicks shortly after their formation. The mass-dependent rate, kick velocity and timing will follow distributions modelled in N-body simulations \cite[e.g.,][]{Oh+2016:2016A&A...590A.107O, Schoettler+2022:2022MNRAS.510.3178S}.

\subsection{Other observables}\label{sec:other-observables}
For high-energy, degenerate, and/or accreting sources formed through binary channels (e.g., X-ray binaries, cataclysmic variables, short gamma-ray bursts, type Ia supernovae), the mapping between binary physical parameters and flux is naturally more complex (and sometimes uncertain) than it is for most stars. This means that predictions for other observables (beyond those current implemented) are more complicated, though not out of reach in many cases. For example, prescriptions for the X-ray luminosity of a given binary exist \citep{Misra+2023}, and we intend to add this feature to \cosmic{} (and thus also to \cogsworth) to make predictions for the X-ray binary populations that have been widely observed with \chandra{} in nearby galaxies. In the future, we will implement mappings for other missions and observables based on their own selection functions.


\section{Conclusions \& Summary}\label{sec:conclusions}

In this paper we have presented \cogsworth, a new open-source code for performing self-consistent population synthesis and galactic dynamics simulations. \cogsworth provides the theoretical infrastructure necessary to make predictions about the positions and velocities of stars and compact objects. We have demonstrated several use cases of the code, showcasing its capabilities to investigate the impact of binary interactions and galactic potentials on the evolution of stars and compact objects - both for intrinsic and observable populations. \cogsworth could be applied to a plethora of investigations on a wide-range of populations, including runaway stars, supernova remnants, X-ray binaries, short gamma-ray bursts and double compact objects.

Given its accessibility and flexibility, we hope that \cogsworth will be a useful tool for the community, enabling and accelerating future studies into binary stars and compact objects.

\section*{Acknowledgements}
We gratefully acknowledge many fruitful discussions with Julianne Dalcanton and Eric Bellm that resulted in several helpful suggestions. TW acknowledges valuable conversations with Matt Orr and Chris Hayward regarding the \fire simulations, and with Alyson Brooks and Akaxia Cruz regarding the \changa simulations. We thank the anonymous referee, Pavel Kroupa and Simon Stevenson for helpful comments on this work. We thank David Hendriks for discussions regarding secondary star ejection velocities. TW thanks the Simons Foundation, Flatiron Institute and Center for Computational Astrophysics for running the pre-doctoral program during which much of this work was completed. The Flatiron Institute is supported by the Simons Foundation. TW and KB acknowledge support from NASA ATP grant 80NSSC24K0768.

\software{\texttt{astropy} \citep{astropy:2013, astropy:2018, astropy:2022}, \texttt{Jupyter} \citep{2007CSE.....9c..21P, kluyver2016jupyter}, \texttt{matplotlib} \citep{Hunter:2007}, \texttt{numpy} \citep{numpy}, \texttt{pandas} \citep{mckinney-proc-scipy-2010, pandas_10957263}, \texttt{python} \citep{python}, \texttt{scipy} \citep{2020SciPy-NMeth, scipy_11702230}, \texttt{Agama} \citep{2019MNRAS.482.1525V}, \texttt{astroquery} \citep{2019AJ....157...98G, astroquery_10799414}, \texttt{ChaNGa} \citep{Jetley2008, Jetley2010, Menon2015}, \texttt{COSMIC} \citep{Breivik2020, COSMIC_13351205}, \texttt{Cython} \citep{cython:2011}, \texttt{dustmaps} \citep{2018JOSS....3..695M, dustmaps_10517733}, \texttt{gaiaunlimited} \citep{Cantat-Gaudin+2023:2023A&A...669A..55C}, \texttt{gala} \citep{gala_JOSS, gala_13377376}, \texttt{h5py} \citep{collette_python_hdf5_2014, h5py_7560547}, \texttt{isochrones} \citep{Morton+2015:2015ascl.soft03010M}, \texttt{legwork} \citep{LEGWORK_joss, LEGWORK_apjs, legwork_12476977}, \texttt{Numba} \citep{numba:2015, Numba_11642058}, \texttt{pynbody} \citep{pynbody, pynbody_10276404}, \texttt{PyTables} \citep{pytables}, \texttt{schwimmbad} \citep{schwimmbad}, \texttt{seaborn} \citep{Waskom2021}, and \texttt{tqdm} \citep{tqdm_3551211}. Some of the results in this paper have been derived using \texttt{healpy} and the HEALPix package\footnote{http://healpix.sourceforge.net} \citep{Zonca2019, 2005ApJ...622..759G, healpy_12746571}.
This research has made use of NASA's Astrophysics Data System. 
Software citation information aggregated using \texttt{\href{https://www.tomwagg.com/software-citation-station/}{The Software Citation Station}} \citep{software-citation-station-paper, software-citation-station-zenodo}.}

\bibliographystyle{aasjournal}
\bibliography{bibs/paper, bibs/software}{}

\restartappendixnumbering

\allowdisplaybreaks
\appendix

\section{Typical simulation code}

In this Section, we demonstrate the code for a typical \cogsworth simulation, to illustrate its ease-of-use and flexibility. The following code block shows how one can run a basic \cogsworth simulation and access and interpret a variety of results.

\usemintedstyle{tango}

\begin{minted}
[
frame=lines,
framesep=2mm,
baselinestretch=1.2,
fontsize=\footnotesize,
linenos
]
{python}
import cogsworth
import gala.potential as gp
import astropy.units as u

# run the simulation
p = cogsworth.pop.Population(
    n_binaries=1000,
    processes=6,
    sfh_model=cogsworth.sfh.Wagg2022,
    galactic_potential=gp.MilkyWayPotential2022(),
    v_dispersion=5 * u.km / u.s,
    max_ev_time=12 * u.Gyr,
    BSE_settings={
        # adjust binary stellar evolution settings here
    },
    sampling_params={
        # adjust initial condition sampling settings here
    }
)
p.create_population()

# access DataFrames of initial conditions + evolution
p.initC, p.bpp

# explore Gala orbits, final positions/velocities
p.orbits, p.final_pos, p.final_vel

# convert to observables (e.g. flux, colour)
p.get_observables(filters=["G", "BP", "RP"],
                  assume_mw_galactocentric=True)

# make some plots
p.plot_cartoon_binary(bin_num=42)
p.plot_orbit(bin_num=42)
p.plot_sky_locations()
cogsworth.plot.plot_cmd(p, "G", "BP", "RP")

# save population for later
p.save("population.h5")
\end{minted}
In relatively few lines of code, this simulation allows users to sample binaries from a SFH, evolve the stars until present day with \cosmic, integrate their orbits through a galactic potential with \gala, convert the intrinsic population to observables and create a series of plots for interpreting the result (including similar plots to Figures~\ref{fig:cartoon-binary} and~\ref{fig:gaia-cmd}). \cogsworth will use the default choices for binary stellar evolution and sampling settings when \texttt{BSE\_settings} and \texttt{sampling\_params} are left empty respectively. Each settings that is individually added to the input dictionary will override the default, such that \texttt{BSE\_settings = \{`alpha1': 0.5\}} would change the efficiency of common-envelope events to 0.5 to leave the other defaults unchanged. For a full list of the settings one case change via \texttt{BSE\_settings} and \texttt{sampling\_params}, see the \href{https://cosmic-popsynth.github.io/}{\cosmic documentation}.

\end{document}